\documentclass{article}

\usepackage{lipsum}
\newcommand\blfootnote[1]{%
  \begingroup
  \renewcommand\thefootnote{}\footnote{#1}%
  \addtocounter{footnote}{-1}%
  \endgroup
}

\usepackage{microtype}
\usepackage{graphicx}
\usepackage{subfigure}
\usepackage{booktabs}
\usepackage{tikz}
\usepackage{float}
\usetikzlibrary{fit, arrows, arrows.meta, shapes.geometric}
\tikzstyle{cnn} = [rectangle, minimum width=3.05cm, minimum height=0.48cm, text centered, draw=black]
\tikzstyle{arrow} = [thick, -{Stealth[scale=1]}]
\definecolor{azul}{RGB}{187, 233, 255}

\usepackage[symbol]{footmisc}
\renewcommand{\thefootnote}{\fnsymbol{footnote}}

\usepackage{hyperref}

\usepackage[accepted]{synsml2023}

\usepackage{amsmath}
\usepackage{amssymb}
\usepackage{mathtools}
\usepackage{amsthm}

\usepackage[capitalize,noabbrev]{cleveref}

\theoremstyle{plain}

\theoremstyle{definition}

\theoremstyle{remark}

\usepackage[textsize=tiny]{todonotes}

\synsmltitlerunning{Learning from Topology: Cosmological Parameter Estimation}

\begin{document}

\twocolumn[
\synsmltitle{Learning from Topology:\\Cosmological Parameter Estimation from the Large-scale Structure}

\synsmlsetsymbol{equal}{*}

\begin{synsmlauthorlist}
\synsmlauthor{Jacky H. T. Yip}{equal,yyy}
\synsmlauthor{Adam Rouhiainen}{equal,yyy}
\synsmlauthor{Gary Shiu}{yyy}
\end{synsmlauthorlist}

\synsmlaffiliation{yyy}{Department of Physics, University of Wisconsin-Madison, Madison, WI 53706, USA}
\synsmlcorrespondingauthor{Jacky H. T. Yip}{hyip2@wisc.edu}

\synsmlkeywords{Cosmological Parameter Estimation, Persistent Homology, Convolutional Neural Networks, Large-scale Structure}

\vskip 0.3in
]

\blfootnote{$^*$Equal contribution. $^1$Department of Physics, University of Wisconsin-Madison, Madison, WI 53706, USA. Correspondence to: Jacky H. T. Yip $<$hyip2@wisc.edu$>$.}

\begin{abstract}
The topology of the large-scale structure of the universe contains valuable information on the underlying cosmological parameters. While persistent homology can extract this topological information, the optimal method for parameter estimation from the tool remains an open question. To address this, we propose a neural network model to map persistence images to cosmological parameters. Through a parameter recovery test, we demonstrate that our model makes accurate and precise estimates, considerably outperforming conventional Bayesian inference approaches.
\end{abstract}

\section{Introduction}
The recent decade has brought the field of cosmology powerful machine learning tools to analyze the vast amount of data from large-scale sky surveys. In this paper, we propose a new method to attack the long-standing problem of cosmological parameter estimation from the large-scale structure of the universe. It has been shown that decent results can be achieved using deep learning (e.g., \citealp{ravanbakhsh17, ntampaka20, wen23, shao23}). However, despite their effectiveness, neural networks trained directly on low-level data (such as dark matter fields and galaxy maps) lack interpretability and provide little understanding of the underlying physics due to the high degrees of freedom in the input data.

Simply put, cosmologists seek summary statistics that not only retain a substantial amount of information but also stem from physical intuitions. In this regard, we consider applying persistent homology (PH) to point clouds of dark matter halos. PH is a topological data analysis tool that quantifies the robustness of topological features across length scales, allowing for a natural description of the multi-scale patterns in the large-scale structure that the halos trace. Recent studies have used this tool to detect primordial non-Gaussianity \cite{biagetti21}, identify cosmic structures \cite{xu19}, and differentiate dark matter models \cite{cisewski-kehe22}.

The raw outputs of a PH computation are persistence diagrams, which can be conveniently vectorized as persistence images. We may flatten and use them directly for Bayesian inference by assuming a Gaussian likelihood via a covariance structure between pixels (e.g., \citealp{cole18}). However, such a high-dimensional covariance matrix often behaves poorly and requires a considerable amount of additional data for its estimation. More critically, the covariance does not capture higher-order correlations, such as patterns spanned by multiple neighbouring pixels.

We propose to train a convolutional neural network (CNN) model on persistence images to estimate the underlying cosmological parameters. For comparisons, we perform the same task by Bayesian inference using two summary statistics: a histogram-based statistic also from PH, and the power spectrum. The former has been recently studied in the literature, while the latter is a baseline statistic ubiquitous in the cosmological community. We find that our CNN model recovers parameters much more accurately and precisely, thus improving on the conventional approaches. This work also serves as a pioneering example of combining computational topology with machine learning in the context of cosmology.

\section{Cosmology and the Large-scale Structure}
The $\Lambda$CDM model is by far the best-supported theory of our universe \cite{dodelson20}. In particular, it explains the evolution of inhomogeneities in the matter distribution in an expanding spacetime. This paper focuses on two parameters in the model: the matter density $\Omega_{\rm{m}}$, and the clustering amplitude $\sigma_8$. While the impacts of these parameters on the evolution are well understood in the linear regime, beyond which we generally resort to N-body simulations. Hence, the inverse problem, parameter estimation from the late-time matter distribution, holds significance both in theory and in practice.

The Zel'dovich approximation, a non-linear model for the evolution of non-interacting particles, predicts that ellipsoidal distributions of matter collapse along their axes into clusters, filaments, and walls \cite{zeldovich70,hidding13}. These components assemble hierarchically to form multi-scale halo clusters, filament loops, and cosmic voids, which are the topological features tracked by PH.

\section{Persistence Statistics}

\subsection{Persistent Homology}
We apply PH to $3$D point clouds of dark matter halos. A point cloud is first triangulated into a simplicial complex, i.e., a set of points, edges, triangles, and tetrahedrons. Then we can define a \emph{parametrized} family of nested (sub)complexes called a filtration. More specifically, as the filtration parameter $\nu$ grows, simplexes are added to the complex according to a designed set of rules. Typically, $\nu$ is interpreted as a length scale, and a simplex is added if its geometric size is less than $\nu$. In this work, we use the $\alpha$DTM$\ell$-filtration \cite{chazal17}, which reduces the disproportionate influence of outliers, and has been shown to be effective for extracting information from the distribution of halos \cite{biagetti21,biagetti22,yip24}.
\begin{figure}[H]
    \centering
    \vspace{-0.15em}
    \centerline{\includegraphics[scale=0.65, trim={0, 0, 0, 0}, clip]{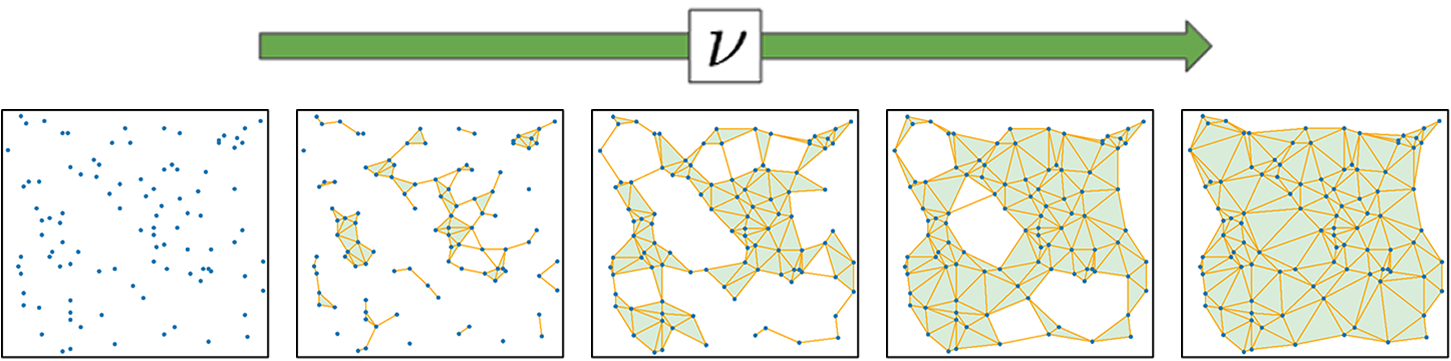}}
    \vspace{-0.15em}
    \caption{Example of a filtration of a $2$D point cloud. The simplicial complex evolves with the filtration parameter $\nu$. Topological features (islands and loops) are created and subsequently trivialized in the process.}
    \label{fig:my_label}
\end{figure}

For every value of $\nu$, we identify the topological features in the corresponding simplicial complex. They are $0$-, $1$-, and $2$-cycles representing islands, closed loops, and enclosed cavities, respectively. They physically correspond to halo clusters, filament loops, and cosmic voids in the large-scale structure. Hence, the evolution of the simplical complex throughout the filtration can be characterized by a set of $(\nu_{\rm{birth}},\nu_{\rm{death}})$ values at which $n$-cycles are born and killed.

\subsection{From Persistence Diagrams to Summary Statistics}
The list of $(\nu_{\rm{birth}},\nu_{\rm{death}})$ pairs can be recast in the $(\nu_{\rm{birth}},\nu_{\rm{persist}}=\nu_{\rm{death}}-\nu_{\rm{birth}})$ coordinates. For each of the three $n$ values, we can plot $(\nu_{\rm{birth}},\nu_{\rm{persist}})$ for all the once-existed $n$-cycles. These three plots called the persistence diagrams are the raw outputs of the PH computation. We can further convert them into useful data vectors:

\emph{Persistence Images} - We assign a Gaussian kernel to each point in a persistence diagram. Then, we discretize the birth-persistence plane and sum up all the kernel densities within each pixel. The $2$D array of numbers obtained is called a persistence image \cite{adam17}.

\emph{Histograms} - Following \citealp{biagetti22} and \citealp{yip24}, we construct histograms from the distributions of $\nu_{\rm{birth}}$ and $\nu_{\rm{persist}}$ for all $n$. The $2\times3=6$ histograms are then concatenated into a $1$D array of numbers.

To apply any canonical machine learning or statistical method, it is necessary to vectorize the observables. Any such vectorization, including the use of persistence images, involves some degree of binning or smoothing and therefore incurs information loss. Nevertheless, the persistence image provides a principled and widely adopted representation that balances fidelity with compatibility. It preserves the density and distributional structure of the persistence diagram while producing a fixed-size, differentiable format well suited for CNN-based models.

\section{The Power Spectrum}
The halo auto power spectrum $P^{\rm{hh}}(k)$ measures the contribution to the overdensity (the difference between the local and mean densities divided by the mean density) of the halo field from modes of wavenumber $k$. It quantifies the ``lumpiness'' of the halo distribution as a function of scale. As a summary statistic, the power spectrum has long been well-understood \cite{peebles80} and used widely to access the cosmological information in the halo field (e.g., \citealp{coulton23-2,jung23}). Hence, it is included in our analysis as a baseline for comparison against the persistence statistics.

We use the persistence images for our CNN model (\textbf{PI-CNN}), and the histograms and power spectrum for Bayesian inference (\textbf{Hist-BI} and \textbf{PS-BI}).
\begin{figure*}
\centering
  \vspace{-0.5em}
  \includegraphics[scale=0.78, trim={0, 0, 0, 0}]{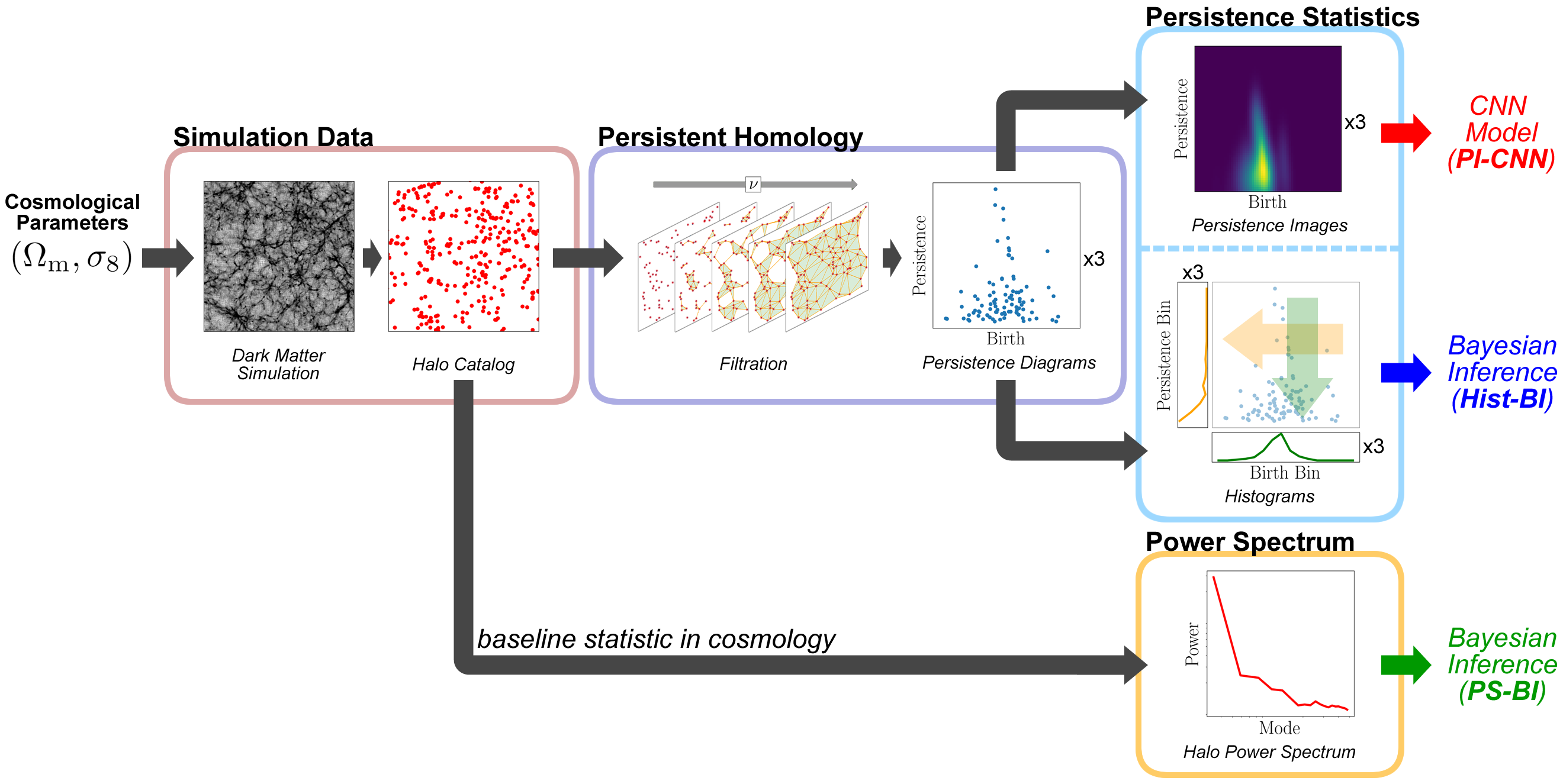}
  \vspace{-1em}
  \caption{Given a $(\Omega_{\rm{m}},\sigma_8)$ pair, we generate a realization of the dark matter field at $z=0.5$ in a $(256\,h^{-1}\rm{Mpc})^3$ box. We locate the halos in the field and apply persistent homology to the halo catalog. The output from the filtration is a list of $(\nu_{\rm{birth}},\nu_{\rm{persist}})$ pairs, plotted as persistence diagrams. We can 1) pixelate the diagrams into persistence images for our CNN model, or 2) sum up the topological features along each axis into histograms for Bayesian inference. ``$\times3$'' refers to having three copies for the $0$-, $1$-, and $2$-cycles. The halo auto power spectrum, also for Bayesian inference, is measured directly from the halo catalog and provides a baseline result for comparison against the persistence statistics.}
\end{figure*}

\section{Simulation and Dataset}
This section details all the computational steps (Figure 2) we take to create our dataset of summary statistics.

All code used in this work, including the dark matter simulation and halo finding setup, the persistent homology and power spectrum pipelines, and scripts for the training of the CNN and Bayesian inference, is now publicly available at \href{https://github.com/jhtyip/TDAflow/tree/main}{github.com/jhtyip/TDAflow/tree/main}. We also provide the key measurements derived from the simulations.

As explained in section $2$, our focus on $\Omega_{\rm m}$ and $\sigma_8$ is primarily motivated by their strong influence on halo clustering. Moreover, the \textsc{FlowPM} simulation code is widely used in the literature for varying these two parameters, which makes our results readily comparable to established benchmarks. While extending the analysis to include additional cosmological parameters is important for broader applicability, doing so would require a substantially larger simulation suite and more complex sampling of a higher-dimensional parameter space, which is beyond the scope of this work. That said, our pipeline is fully general and can be extended to incorporate additional parameters, given sufficient training data and computational resources.

\subsection{Dark Matter Simulation}
We employ \textsc{FlowPM} \cite{modi21} to produce snapshots of late-time distributions of dark matter particles. Although it lacks exact dynamics, the fast solver is sufficiently accurate for this work as a proof of concept (see \citealp{feng16} for discussions on accuracy issues). Besides, stability theorems ensure that PH outputs are robust to noise in the input data under certain assumptions \cite{turkes21}.

Given a pair of cosmological parameter values, we can compute the primordial matter power spectrum, from which a Gaussian random field at a required redshift can be generated via the transfer function \cite{eisenstein98} and linear growth factors. We begin evolving the Gaussian field at $z=9$ for numerical stability, and a total of $10$ time-steps are used. The snapshot is taken at redshift $z=0.5$ to match standard survey samples (e.g., the BOSS CMASS galaxies, \citealp{reid16}) for the convenience of future analyses on galaxy maps. We note that our pipeline can be applied to data at other redshifts without modification. Each simulation box is $(256\,h^{-1}\rm{Mpc})^3$ in volume and contains $(160)^3$ particles, resulting in a particle resolution on par with state-of-the-art simulations such as the \textsc{Quijote} suite \cite{vn20}.

For both the training of our CNN model and the likelihood evaluation, we vary $\Omega_{\rm{m}}$ and $\sigma_8$ in our simulations, where $\Omega_{\rm{m}}\in[0.21,0.41]$ and $\sigma_8\in[0.72,0.92]$. Each range is divided into $60$ equal intervals for a total of $3600$ distinct $(\Omega_{\rm{m}},\sigma_8)$ configurations, and we generate $10$ independent realizations for each configuration for averaging. In our cosmological setting, averaging over realizations with the same parameters is equivalent to increasing the observed survey volume, which reduces sample variance and yields more stable statistics. Averaging itself does not introduce overfitting risk, since each averaged statistic is associated with a distinct cosmological model. We also note that the volume of each realization is relatively small, so averaging over $10$ realizations corresponds to an effective volume that remains well within the range of modern surveys. In summary, there is a grand total of $36000$ simulations in the main dataset. We use the Planck 2015 results for all other cosmological parameters (final column of Table 4 in \citealp{planck16}). Moreover, a flat universe is assumed such that we take the dark energy density as $\Omega_\Lambda=1-\Omega_{\rm{m}}$ whenever applicable.

We further produce $15000$ realizations at the fiducial cosmology $(\Omega_{\rm{m}},\sigma_8)=(0.3089,0.8159)$. We use $5000$ of these to estimate the covariance matrix in the likelihood and the remaining $10000$ for the parameter recovery test.

\subsection{Halo Catalog}
We use the \textsc{rockstar} halo finder \cite{behroozi13} to identify dark matter halos in the simulation snapshots. We set the force resolution to $0.005\,h^{-1}\rm{Mpc}$ and keep other settings as default. In each fiducial box, $\rm{\sim}2000$ halos are identified, which is comparable to \textsc{Quijote} as a credibility check. Hence, each halo catalog is a list of halo positions in real space to which we apply PH.

\subsection{Summary Statistics}
Our PH code relies on the \textsc{GUDHI} library \cite{gudhi15}. From persistence diagrams to images, we employ \textsc{scikit-learn} to fit the Gaussian kernel density model. Each point in the diagrams is also weighted by $\sqrt{\nu_{\rm{persist}}}$ to emphasize more persistent features. There are two key parameters in generating the persistence images: the kernel bandwidth and the image resolution. We perform a grid search over the combinations \(\{1, 2, 3, 5\} \times \{32^2, 64^2, 128^2, 256^2\}\) and select the setting \((2, 64^2)\), which balances performance with memory efficiency. In general, a larger kernel bandwidth should be used for nosier and more sparse persistence diagrams. For the histograms, we use the same bins across all parameter configurations. After concatenation, the full histogram statistic is further downsampled to $6\times16=96$ numbers for optimal inference accuracy.

We use \textsc{Pylians} \cite{Pylians18} to compute the halo auto power spectrum. There is only the monopole since the halos are in real space. The power spectrum is truncated by imposing $k_{\rm{max}}=0.4\,h\rm{Mpc}^{-1}$, for the reason that our halo catalogs from the fast simulations might not be very reliable in the highly non-linear regime (we checked that the inference results are similar for $k_{\rm{max}}=0.3$, $0.4$, and $0.5\,h\rm{Mpc}^{-1}$, so our conclusion does not depend on this choice). We also model the shot noise as Poisson distributed and subtract $1/n_{\rm{h}}$ from the power spectrum, where $n_{\rm{h}}$ is the catalog's halo number density. However, for the estimation of the covariance matrix, the shot noise is kept because it contributes as a source of uncertainty.

\section{Estimation Methods}

\subsection{Convolutional Neural Network Model}
We design a neural network model that combines in parallel a CNN with a stack of dense layers to map the persistence images to $(\Omega_{\rm{m}},\sigma_8)$. The inputs of the model are the $0$-, $1$-, and $2$-cycle persistence images stacked into $3$ channels, and the outputs are $2$ numbers for the two parameters (Figure 3).

The CNN side of our parallel networks has four sequential blocks, each made of a $3\times3$ convolution with no padding, and ReLU activation functions. Each convolution is followed by a $2\times2$ max pooling layer with stride $2$. After the fourth block, the data is reduced to $2^2$ pixels, and two dense layers with ReLUs follow to output $2$ numbers.

On the dense side, we turn the persistence images into a $1$D array by summing along the birth and persistence axes of the images. As this summed data contains only $3\times2\times64=384$ numbers, we use a stack of dense layers on the entire data without pooling. We tested different combinations of $1$D convolutions and pooling, with and without ResNet blocks, and have found no increase in model performance over simply using a stack of five dense layers with ReLUs.

We find a modest increase in inference precision (variances are reduced by $\rm{\sim}10\%$) with the parallel networks over the CNN side alone. The $2$ numbers from each side of the model are finally averaged into our $(\Omega_{\rm{m}},\sigma_8)$ estimate. 

We use $33000$ persistence images for the training set and $3000$ for the validation set. For every $(\Omega_\text{m},\sigma_8)$ configuration, we average $10$ corresponding images pixelwise. While this averaging does reduce the number of images for training by a factor of $10$, we find increased precision of our model. Our loss function is the mean squared error between the model outputs and the true parameter values. By analyzing the training and validation performance, we have carefully chosen the number of parameters in our architecture (detailed in Figure 3) to maximize performance while preventing overfitting. Our model has $1.34$ million parameters in total. To further help prevent overfitting, we use a small batch size of $16$. We train with a learning rate of $10^{-4}$ with the Adam optimizer, decreased by a factor of $0.75$ when the loss plateaus.

A potential hurdle we suspect in using a traditional CNN on persistence images is the overall smoothness of each image. In a traditional image classification dataset, we encounter, e.g., many distinct edges in each image; the convolutional filters in the first few layers of the network would learn to detect these edges. It may require a more novel approach to capture more of the features of persistence images.

\begin{figure}
\begin{center}

\begin{tikzpicture}[node distance=0.58cm]

\scriptsize

\pgfdeclarelayer{bg}
\pgfsetlayers{bg, main}

\hspace{0.2cm}\node (start) [cnn] {Persistence images: $3\times64^2$};
\node (cnn0) [cnn, below left=0.1cm and -4.2em of start, fill=azul!75] {$3\times3$ conv: $64\times62^2$};
\node (cnn1) [cnn, below of=cnn0, fill=red!15] {$2\times2$ max pool: $64\times31^2$};
\node (cnn2) [cnn, below of=cnn1, fill=azul!75] {$3\times3$ conv: $64\times29^2$};
\node (cnn3) [cnn, below of=cnn2, fill=red!15] {$2\times2$ max pool: $64\times14^2$};
\node (cnn4) [cnn, below of=cnn3, fill=azul!75] {$3\times3$ conv: $128\times12^2$};
\node (cnn5) [cnn, below of=cnn4, fill=red!15] {$2\times2$ max pool: $128\times6^2$};
\node (cnn6) [cnn, below of=cnn5, fill=azul!75] {$3\times3$ conv: $128\times4^2$};
\node (cnn7) [cnn, below of=cnn6, fill=red!15] {$2\times2$ max pool: $128\times2^2$};
\node (cnn8) [cnn, below of=cnn7, fill=green!15] {Dense: $128$};
\node (cnn9) [cnn, below of=cnn8, fill=green!15] {Dense: $2$};

\node (nn0) [cnn, below right=5.13em and -4.2em of start, fill=blue!0, node distance=0.89cm] {Sum along $x,y$: $3\times2\times64$};
\node (nn1) [cnn, below of=nn0, fill=green!15] {Dense: 512};
\node (nn2) [cnn, below of=nn1, fill=green!15] {Dense: 1024};
\node (nn3) [cnn, below of=nn2, fill=green!15] {Dense: 256};
\node (nn4) [cnn, below of=nn3, fill=green!15] {Dense: 128};
\node (nn5) [cnn, below of=nn4, fill=green!15] {Dense: 2};

\node (end) [cnn, below right=0.1cm and -4.2em of cnn9, node distance=0.89cm] {$(\Omega_{\rm{m}},\sigma_{8})$};

\draw [arrow] (start.south) to (cnn0.east);
\draw [arrow] (start.south) to (nn0.north);

\draw [arrow] (cnn0.east) to [out=-30,in=30](cnn1.east) node[yshift=0.3cm] {\hspace{1.1cm}ReLU};
\draw [arrow] (cnn1.east) to [out=-30,in=30](cnn2.east);
\draw [arrow] (cnn2.east) to [out=-30,in=30](cnn3.east) node[yshift=0.3cm] {\hspace{1.1cm}ReLU};
\draw [arrow] (cnn3.east) to [out=-30,in=30](cnn4.east);
\draw [arrow] (cnn4.east) to [out=-30,in=30](cnn5.east) node[yshift=0.3cm] {\hspace{1.1cm}ReLU};
\draw [arrow] (cnn5.east) to [out=-30,in=30](cnn6.east);
\draw [arrow] (cnn6.east) to [out=-30,in=30](cnn7.east) node[yshift=0.3cm] {\hspace{1.1cm}ReLU};
\draw [arrow] (cnn7.east) to [out=-30,in=30](cnn8.east) node[yshift=0.3cm] {\hspace{1.175cm}Flatten};
\draw [arrow] (cnn8.east) to [out=-30,in=30](cnn9.east) node[yshift=0.3cm] {\hspace{1.1cm}ReLU};

\draw [arrow] (nn0.east) to [out=-30,in=30](nn1.east) node[yshift=0.3cm] {\hspace{1.175cm}Flatten};
\draw [arrow] (nn1.east) to [out=-30,in=30](nn2.east) node[yshift=0.3cm] {\hspace{1.1cm}ReLU};
\draw [arrow] (nn2.east) to [out=-30,in=30](nn3.east) node[yshift=0.3cm] {\hspace{1.1cm}ReLU};
\draw [arrow] (nn3.east) to [out=-30,in=30](nn4.east) node[yshift=0.3cm] {\hspace{1.1cm}ReLU};
\draw [arrow] (nn4.east) to [out=-30,in=30](nn5.east) node[yshift=0.3cm] {\hspace{1.1cm}ReLU};

\draw [arrow] (cnn9.east) to (end.north);
\draw [arrow] (nn5.south) to (end.north);

\end{tikzpicture}

\vspace{0em}

\caption{Architecture of our CNN model, with layer names and output dimensions.}

\end{center}
\label{fig:CNN}
\end{figure}
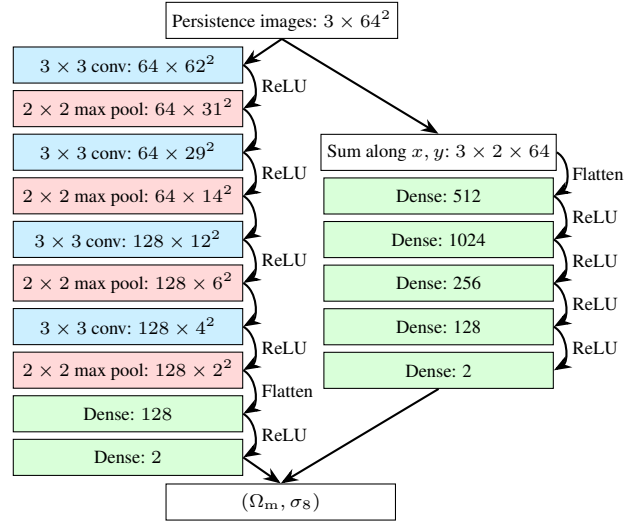

\subsection{Bayesian Inference}
We adopt a Gaussian likelihood $\mathcal{L}(\boldsymbol{\theta}|\boldsymbol{D})$ and a flat prior $p(\boldsymbol{\theta})$ such that the log-posterior is
\begin{align*}
\begin{split}
    \ln p(\boldsymbol{\theta}|\boldsymbol{D})=-\frac{1}{2}(\boldsymbol{D}-\boldsymbol{\mu}(\boldsymbol{\theta}))^TC^{-1}(\boldsymbol{D}-\boldsymbol{\mu}(\boldsymbol{\theta}))+\rm{const.}
\end{split}
\end{align*}
by Bayes' theorem. Here $\boldsymbol{\theta}$ is one of the $3600$ $(\Omega_{\rm{m}},\sigma_8)$ configurations at which the log-posterior is to be numerically evaluated. $\boldsymbol{\mu}(\boldsymbol{\theta})$ is the data vector (the full histogram statistic or the power spectrum) measured at $\boldsymbol{\theta}$ and averaged over $10$ realizations. $\boldsymbol{D}$ is an observation, also averaged over $10$ realizations, measured at some unknown configuration that we want to recover. $C$ is the covariance matrix, and we include the Hartlap factor (Eq. (17) in \citealp{hartlap06}) for the unbiased estimation of $C^{-1}$. The constant term ${\rm{const.}}=\ln\frac{p(\boldsymbol{\theta})}{p(\boldsymbol{D})}$ can be ignored as it plays no role in finding the maximum a posteriori (MAP) estimate.

It is worth noting that the likelihood function of the histogram statistic may be checked empirically to be Gaussian \cite{biagetti22,yip24}. For the power spectrum, assuming a Gaussian likelihood is customary \cite{carron13}. By additionally averaging over realizations (such that the central limit theorem helps), the Gaussian likelihood for inference is well-motivated overall.

We use a $2$D Gaussian filter to smooth out the noise in the numerically evaluated log-posterior, with $\sigma$ of the kernel set to $1$ and $0.2$ pixels (optimized for inference accuracy) for the histogram statistic and the power spectrum, respectively. The MAP estimate is the $(\Omega_{\rm{m}},\sigma_8)$ configuration that maximizes the (log-)posterior.

\section{Results and Discussion}

We conduct a parameter recovery test to compare the PI-CNN, Hist-BI, and PS-BI pipelines. All statistics are averaged over every $10$ boxes in the test dataset of $10000$ fiducial boxes, i.e., we have $1000$ independent observations at $(\Omega_{\rm{m}},\sigma_8)=(0.3089,0.8159)$. This averaging is in accordance with the estimation methods, as described in the previous section.

While in this test we focus on recovering Planck fiducial parameter values as a physically well-motivated benchmark, we reiterate that the CNN is trained on simulations spanning a broad and representative range of $(\Omega_{\rm m}, \sigma_8)$ values. The model therefore learns from a diverse set of input-output mappings, rather than being specialized to any single point in the parameter space. We do not expect the architecture to be fine-tuned to perform well only at this point.

We present the results in Figure 4. In the central panel, we plot the PI-CNN estimates in red, Hist-BI in blue, and PS-BI in green. The MAP estimates are restricted to grid points because the log-posterior is evaluated on the discretized $\Omega_{\rm{m}}$-$\sigma_8$ plane. We fit a bivariate Gaussian to each of the distributions and plot the contours for the $68\%$ and $95\%$ confidence regions. The marginalized Gaussians are shown in the side panels. The mean estimates are $(0.3088\pm0.0105,0.8160\pm0.0154)$ for PI-CNN, $(0.2903\pm0.0256,0.8259\pm0.0191)$ for Hist-BI, and $(0.3294\pm0.0464,0.8320\pm0.0329)$ for PS-BI. The true fiducial values are marked by the crosshairs.

\begin{figure}
    \centering
    \centerline{\includegraphics[scale=0.6, trim={0, 0, 3em, 5em}, clip]{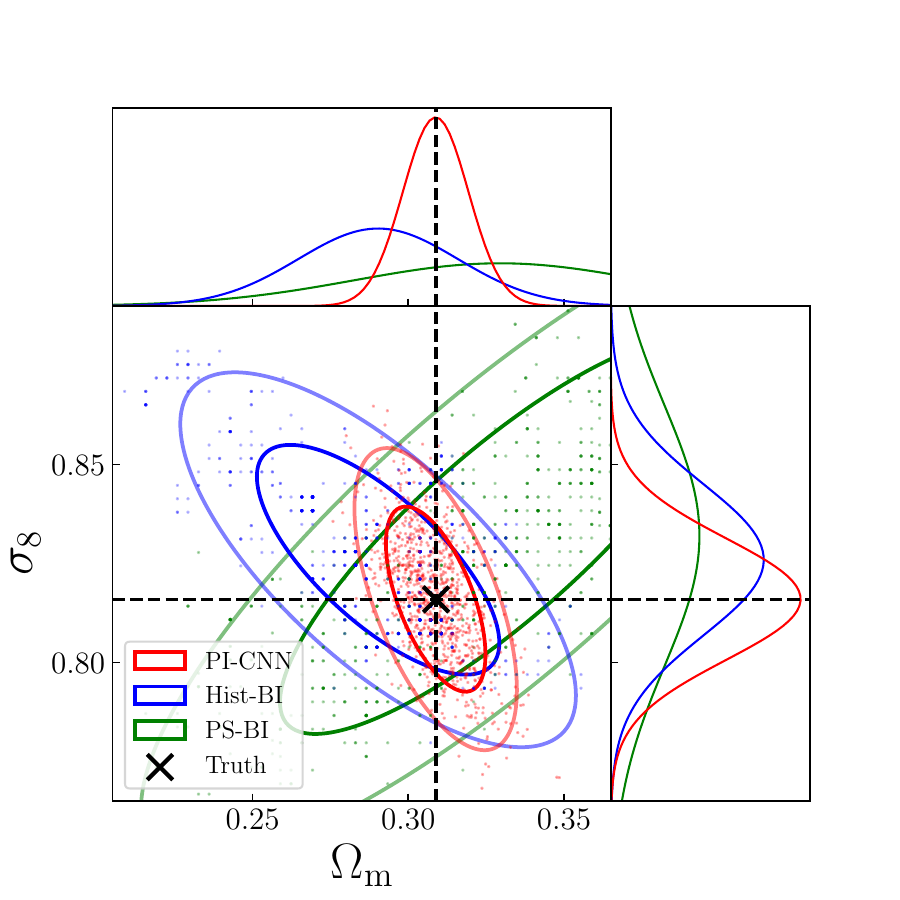}}
    \vspace{-0.5em}
    \caption{Red, blue, and green dots in the central panel mark the $(\Omega_{\rm{m}},\sigma_8)$ values estimated on the $1000$ fiducial measurements by the CNN model and the Bayesian inference approaches. We fit a Gaussian to each distribution; the contours mark the $68\%$ and $95\%$ confidence regions. Side panels display the marginalized Gaussians. The crosshairs locate the true values.}
    \label{fig:my_label}
\end{figure}

Comparing between the PH-based methods, the PI-CNN Gaussian is accurately centered on the true value, while the Hist-BI Gaussian is fairly biased. The variances of the PI-CNN estimates are also lower, particularly for $\Omega_{\rm{m}}$. Both methods exhibit a negative correlation between the two parameters, which is a common result for many summary statistics (e.g., the mass function of low-redshift galaxy clusters, \citealp{vikhlinin09}). Intriguingly, the Pearson correlation coefficient for PI-CNN ($-0.60$) is less negative than that for Hist-BI ($-0.72$). This implies that the $2$D persistence images contain extra information extractable by our CNN model for breaking partially the parameter degeneracy.

The PS-BI result is the most biased and has much higher variances. This is not surprising because the power spectrum is a two-point statistic, while PH theoretically probes higher-order correlations as well, considering that topological features are non-local and are each constructed from many points typically (though arguably PH misses out some long-range information). Interested readers may refer to \citealp{yip24} for a comprehensive analysis on the cosmological information content from PH compared with the two- and three-point statistics.

We note that although the PI-CNN approach yields more constrained contours, they are not systematically offset from those obtained using the other two methods and show no statistical inconsistencies. Hence, we do not expect this approach, on its own, to resolve phenomena such as the $\sigma_8$ tension.

To investigate which topological features contribute most to the predictive power of the model, and to gain insight into what the CNN is learning from, we conduct an indirect analysis by training separate CNN models on persistence images corresponding to $0$-, $1$-, and $2$-cycles. Each model uses the same architecture and training setup as the full PI-CNN approach, with the only modification being an adjustment of the input layer in each side of the parallel networks to match the single homological dimension input.

For these additional models, we present the parameter recovery test results in Figure 5, showing contours marking the $68\%$ confidence regions. We find that all models exhibit a negative correlation between the two parameters, consistent with the full model. This indicates that topological features across homological dimensions are sensitive to the parameters in a similar manner. Among the three dimensions, the result from the $0$-cycles provides the most constraining contour, but shows a considerable bias in the recovered values of $\Omega_{\rm m}$. As the full model yields nearly unbiased estimates, the bias must be mitigated by information contained in the $1$- and $2$-cycles. The $0$-cycles being the most informative is expected, as their birth values from the $\alpha$DTM$\ell$-filtration directly encode the local halo density around each halo, which is strongly influenced by $\Omega_{\rm m}$ and especially by $\sigma_8$.

\begin{figure}
    \centering
    \centerline{\includegraphics[scale=0.6, trim={0, 0, 3em, 5em}, clip]{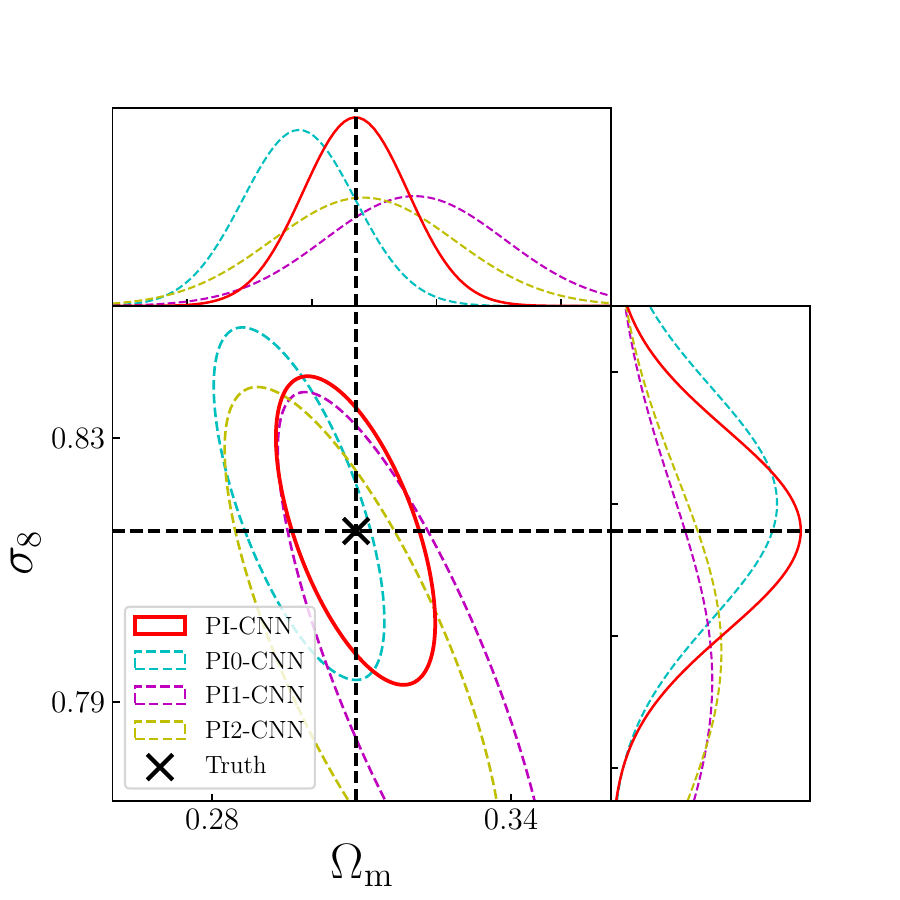}}
    \vspace{-0.5em}
    \caption{Cyan, magenta, and yellow contours mark the $68\%$ confidence regions from $(\Omega_{\rm{m}},\sigma_8)$ values estimated on the $1000$ fiducial measurements by the CNN models trained on persistence images corresponding to $0$-, $1$-, and $2$-cycles, respectively. Similarly, the red contour, which is the same as the one shown in Figure 4, is from the full PI-CNN approach. Side panels display the marginalized Gaussians. The crosshairs locate the true values.}
    \label{fig:my_label}
\end{figure}

While the current results are based entirely on simulated halo catalogs, our pipeline is designed to be compatible with real observational data. In particular, we selected a redshift range that aligns with ongoing galaxy surveys, which trace the underlying halo distribution. Applying the method to real data is an important next step, which will involve additional challenges such as survey geometry, selection effects, and redshift uncertainties, which are beyond the scope of this feasibility study.

\section{Conclusion}
We have put cosmology, computational topology, and machine learning together by training a CNN model on persistence images for cosmological parameter estimation. In the parameter recovery test, our CNN model made accurate and precise estimates, outperforming Bayesian inference methods that use a histogram-based persistence statistic and the power spectrum respectively. We have also observed that the persistence statistics outperform the power spectrum regardless of the estimation method, implying that PH can effectively extract more cosmological information than the two-point statistic.

\section*{Acknowledgements}
The authors declare no conflict of interest. The authors thank Moritz Münchmeyer and Matteo Biagetti for useful discussions. This material is based upon work supported by the U.S. Department of Energy, Office of Science, Office of High Energy Physics under Award Numbers DE-SC-0023719 and DE-SC-0017647. The dark matter simulations, persistent homology computations, and power spectrum measurements were conducted using CHTC resources \cite{CHTC06}.

\bibliography{example_paper}
\bibliographystyle{synsml2023}

\end{document}